\documentclass[12pt]{iopart}
\usepackage{iopams, graphics, epsfig}

\def\be{\begin{equation}}
\def\ee{\end{equation}}

\begin{document}

\title{Reply to `A comment on `The Cauchy problem of $f(R)$ 
gravity''}

\author{Valerio Faraoni}

\address{Physics Department, Bishop's University, 2600 College St.,
Sherbrooke, Qu\'{e}bec, Canada J1M~1Z7}

\eads{\mailto{vfaraoni@ubishops.ca}}
\date{}

\begin{abstract}
We reply to a comment by Capozziello and Vignolo
about the Cauchy problem of Palatini $f(R)$ gravity.
\end{abstract}


\vskip0.5truecm
In \cite{CV} Capozziello and Vignolo comment on 
Ref.~\cite{Cauchy}, which 
discusses the Cauchy problem of metric and Palatini $f(R)$ 
gravity (see also \cite{CV2}). The contentious issue is Palatini 
$f(R)$ gravity {\em in 
vacuo}. Ref.~\cite{Cauchy} reduces the initial value problem 
of 
both 
metric and Palatini $f(R)$ gravity to the corresponding one for 
$\omega_0=0$ and $\omega_0=-3/2$ Brans-Dicke theory, 
respectively, 
using a well-known equivalence theorem.  A $3+1$ ADM 
decomposition is performed, matter quantities $E^{(m)}, S^{(m)}$, 
and $ { S^{(m)}_i}^j$ are defined, and the dynamical equations 
are reduced to
\begin{eqnarray}
&& \partial_t {K^i}_j + N^l\partial_l {K^i}_j + 
{K^i}_l\partial_j N^l - 
{K_j}^l\partial_l  N^i 
+ D^i D_j N \nonumber \\
&&\nonumber \\
&& - ^{(3)}{R^i}_j N - NK{K^i}_j + \frac{N}{2\phi}\delta^i_j 
\left( 2V(\phi) + \square
\phi \right) + \frac{N}{\phi} \left( D^i Q_j + \Pi {K^i}_j  
\right) \nonumber 
\\
&&\nonumber \\
&& +  \frac{N\omega_0}{\phi^2}Q^i Q_j = \frac{N}{2\phi}\left( 
\left( S^{(m)
} - E^{(m)} \right) \delta^i_j - 2 {S^{(m) \,\, i}}_j \right) 
\;,\label{43}\\
&&\nonumber \\
&&  \partial_t K + N^l\partial_l K + ^{(3)}\Delta N - 
NK_{ij}K^{ij} - \frac{N
}{\phi} \left( D^{c}Q_{c} + \Pi K \right) - 
\frac{\omega_{0}N}{\phi^2}\Pi^2 \nonumber \\
&&\nonumber \\
 &&= \frac{N}{2\phi}\left[ -2V(\phi) - 3\square\phi + S^{(m)} + E^{(m)} 
\right] \;,\label{44}
\end{eqnarray}
and 
\begin{equation}
\label{45}
\left( \omega_0 + \frac{3}{2} \right)\square\phi = 
\frac{T^{(m)}}{2} - 
2V(\phi) + \phi V^{\prime}(\phi) + 
\mathbf{\frac{\omega_{0}}{\phi}}\left( \Pi^{2}
 - Q^{2} \right) 
\end{equation}
(eqs.~(4.3)-(4.5) of \cite{Cauchy}, see this reference for 
details and 
precise definitions of the quantities appearing here). In 
\cite{Cauchy} it was stated that, in the $\omega_0=-3/2$ 
Brans-Dicke equivalent of Palatini $f(R)$ gravity, the dynamical 
equation~(\ref{45}) for $\phi$ disappears and, therefore,  it is 
impossible to 
eliminate $\Box \phi$ from eqs.~(\ref{43}) and (\ref{44}) (we 
remind the reader that the 
task is to eliminate all the second derivatives of the ADM 
variables to reduce the system to a fully first order form, and 
that  
these second derivatives appear only through  $\Box \phi$, as 
is clear from the inspection of eqs.~(\ref{43}) and 
(\ref{44})).   
Capozziello and Vignolo point out that, {\em in vacuo}, the 
scalar 
$\phi$ must satisfy  the equation
\be\label{2}
f'(\tilde{R})\tilde{R}-2f(\tilde{R})=0 \;,
\ee
where $\tilde{R}\equiv g^{ab}\tilde{R}_{ab}$ is the Ricci scalar
built out of the non-metric connection of Palatini theory 
(beware of different notations, contrary to \cite{CV} we reserve 
the symbol $R$ for the Ricci scalar built with the metric 
connection of $g_{ab}$). As a consequence of eq.~(\ref{2}), 
$\tilde{R}$ and, consequently, $\phi \equiv f'(\tilde{R})$ are 
constant and $\Box \phi$ vanishes identically, leaving no second 
derivatives of $\phi$ in eqs.~(\ref{43}) and (\ref{44}) and 
a well-formulated (and, eventually, well-posed) Cauchy 
problem for vacuum Palatini $f(R)$ gravity.  Capozziello and 
Vignolo correctly note that  Palatini modified gravity {\em in 
vacuo} 
reduces to 
General Relativity with a cosmological constant, which is 
well-known to have a well-posed Cauchy problem and therefore the 
statement in \cite{Cauchy} is inconsistent with this 
case.\footnote{The fact that Palatini $f(R)$ gravity {\em in 
vacuo} 
reduces to General Relativity with a cosmological constant is also 
noted in \cite{OlmoSingh} and was overlooked in \cite{Cauchy}.} 

We fully agree with the argument of Capozziello and 
Vignolo (the essence of this argument was addressed, at the same time, in 
the review paper~\cite{review}, to which we refer the reader for further details).  
In  fact, their argument can be extended a little beyond the 
vacuum 
case. The scalar $\tilde{R}$ always satisfies the equation
\be\label{3}
f'(\tilde{R})\tilde{R}-2f(\tilde{R})=8\pi G \,T 
\ee
(where $T$ is the trace of the matter energy-momentum tensor) 
obtained by taking the trace of the field equations. In {\em any} 
situation in which $T$ is constant, including as special cases 
vacuum and conformal matter (for which $T=0$), $\tilde{R}$ and 
$\phi$ are constant and $\Box \phi$ vanishes identically, 
dropping  out of eqs.~(\ref{43}) and (\ref{44}) and leaving a 
well-posed 
Cauchy problem. The calculations of \cite{Cauchy} are correct but 
the statement that the Cauchy problem is always ill-formulated 
for Palatini modified gravity appearing there is 
definitely incorrect. Although it is mentioned in \cite{Cauchy} 
that the Cauchy problem will still be well-formulated when $\Box 
\phi=0$ and in particular when $\phi=$~constant, the significance 
of this situation and the fact that it comprises vacuum and 
$T=$~const. were missed. 

What about the situation $T\neq$~constant in Palatini $f(R)$ 
gravity? In this case eq.~(\ref{3}) implies that $\phi$ can be 
expressed as  a function of $T$ and, in eqs.~(\ref{43}) and 
(\ref{44}), $\Box\phi$ is replaced by $ \frac{d\phi}{dT}\, 
\Box T+ \frac{d^2\phi}{dT^2}\, \nabla^c T\nabla_c T$. It seems 
virtually impossible to eliminate time derivatives and all  
second derivatives  of $T$ from these equations: it would be 
necessary to express all these derivatives as functions of first 
space derivatives of $T$ or of the other variables. Therefore, 
the 
Cauchy problem in this case is likely to be ill-formulated. This is not a 
rigorous theorem at this stage, but it seems that only extremely 
contrived forms of matter could satisfy this stringent 
mathematical requirement.\footnote{In addition, imposing that 
eq.~(\ref{3}) has real solutions constrains the admissible forms 
of the function $f(\tilde{R})$. Here we assume that such 
solutions exist.} The recent interest in Palatini $f(R)$ gravity 
is due to the need of explaining the present acceleration of the 
universe without dark energy \cite{Vollick}. For  a general 
Friedmann-Lemaitre-Robertson-Walker universe experiencing a 
sequence of inflationary, matter, radiation, and accelerated 
eras, the cases of vacuum or $T=$~constant are clearly irrelevant. 
However, the point of Capozziello and Vignolo stands for the 
situations discussed.

\section*{Acknowledgements}

It is a pleasure to thank  Thomas Sotiriou for an illuminating 
discussion and for comments, and a Bishop's University Research Grant and the 
Natural Science and  Engineering Research Council of Canada for 
financial support.

\end{document}